# Spin transistor built on 2D van der Waals heterostructures

Shengwei Jiang[1,2], Lizhong Li[2], Zefang Wang[2,3], Jie Shan[1,2,4*], and Kin Fai Mak[1,2,4*]

[1]Laboratory of Atomic and Solid State Physics, Cornell University, Ithaca, NY, USA
[2]School of Applied and Engineering Physics, Cornell University, Ithaca, NY, USA
[3]Department of Physics, Penn State University, University Park, PA, USA
[4]Kavli Institute at Cornell for Nanoscale Science, Ithaca, NY, USA

[*]E-mails: kinfai.mak@cornell.edu; jie.shan@cornell.edu.

**Spin transistors (whose on-off operation is achieved by electric-field-controlled spin orientation [1]), if realized, can revolutionize modern electronics through the implementation of a faster and a more energy-efficient performance as well as non-volatile data storage [2,3]. The original proposal by Datta and Das [1] that relies on electric-field-controlled spin precession in a semiconductor channel faces significant challenges including inefficient spin injection, spin relaxation and spread of the spin precession angle [4,5]. Recent demonstration of electric-field switching of magnetic order [6-8] and spin filtering [9-12] in two-dimensional magnetic insulator $CrI_3$ has inspired a new operational principle for spin transistors. Here we demonstrate spin field-effect transistors based on dual-gated graphene/$CrI_3$ tunnel junctions. These devices show an ambipolar transistor behavior and tunnel magnetoresistance widely tunable by gating when the $CrI_3$ magnetic tunnel barrier undergoes an antiferromagnetic-ferromagnetic spin-flip transition. Under a constant magnetic bias in the vicinity of the spin-flip transition, the gate voltage can repeatedly alter the device between a high and a low conductance state with a large hysteresis. This new spin transistor concept based on the electric-field-controlled spin-flip transition in the magnetic tunnel barrier is immune to interface imperfections and allows spin injection, control and detection in a single device.**

The discovery of two-dimensional (2D) magnetic crystals [13-22] has provided a unique platform to explore new spintronic device concepts with unprecedented controllability. $CrI_3$, a layered magnetic insulator [23], has shown an intriguing layer thickness-dependent magnetic order [13]: whereas monolayer $CrI_3$ is an Ising ferromagnet with spins pointing out-of-plane, bilayer $CrI_3$ is an antiferromagnet consisting of two ferromagnetic monolayers with antiparallel spins. Under a magnetic field of less than 1 T, bilayer $CrI_3$ can be turned into a ferromagnet through a spin-flip transition, suggesting an interlayer exchange interaction on the order of ~ 0.2 meV [7]. Such a weak interlayer interaction opens the possibility of inducing the spin-flip transition by means other than a magnetic field. Indeed, recent experiments have demonstrated antiferromagnet-ferromagnet switching in bilayer $CrI_3$ by either a pure electric field [6] or electrostatic doping [7,8] under a constant magnetic bias. Another parallel recent advance in atomically thin magnetic crystals is the discovery of large tunnel magnetoresistance (TMR) in magnetic tunnel junctions utilizing 2D $CrI_3$ as a tunnel barrier [9-12]. Because of the efficient spin filtering effect in 2D $CrI_3$, a change in the tunnel resistance by ~ 100% has been observed when bilayer CrI$_3$ undergoes an antiferromagnet-ferromagnet transition driven by an external magnetic field. Magnetoresistance thus provides an effective means to read the magnetic state of 2D CrI$_3$. These two important results, when combined, provide the basis for a conceptually new spin transistor that we investigate in this study.

Our device consists of a van der Waals heterostructure of bilayer graphene/bilayer CrI$_3$/bilayer graphene with top and bottom gates (Fig. 1a). See Methods for fabrication details and Supplementary Sections 1 and 2 for basic materials and device characterization. The vertical tunnel junction utilizes bilayer graphene as the source and drain contacts to achieve an ambipolar transistor as we demonstrate below, and bilayer CrI$_3$, as the magnetic tunnel barrier. The two nearly symmetric gates are made of few-layer graphene gate electrodes and hexagonal boron nitride (hBN) gate dielectrics of ~ 20 nm. The gates tune the Fermi level of the nearest bilayer graphene contact effectively, which modulates the conductance of the junction through resonant tunneling [24, 25]. This effectively is a tunnel field-effect transistor (TFET). The gates can also cause a significant modulation in the tunnel conductance by inducing a spin-flip transition in bilayer CrI$_3$ [6-8]. A spin-TFET action is thus realized through electrical switching of spins in the magnetic tunnel barrier (Fig. 1a). This is in contrast to the original spin field-effect transistor proposed by Datta and Das that relies on the gate-controlled spin precession in a semiconductor channel [1].

Figure 1b is the tunnel conductance $G$ measured under a small bias (4 mV) as a function of top gate voltage $V_{tg}$ in the absence of back gate voltage and external magnetic field. The measurement was performed at 4 K, at which bilayer CrI$_3$ is an antiferromagnet. Measurements at varying temperatures are included in Supplementary Sect. 3. Here an ambipolar transistor behavior with a zero conductance (*i.e.* off) state is observed for graphene tunnel junctions. This is a consequence of a sizable band gap opened in both bilayer graphene contacts [26, 27] by a build-in interfacial electric field from the asymmetric hBN/bilayer graphene/bilayer CrI$_3$ structure (no off state is observed when gapless monolayer graphene is used as contacts, see Supplementary Sect. 6). A detailed bias dependence measurement, combined with modeling of resonant tunneling in the heterostructure, shows that the gap is around 100 meV (see Supplementary Sect. 5). The result also shows that both bilayer graphene contacts are unintentionally *p*-doped (likely due to charge transfer to CrI$_3$). Since the top gate is effectively screened by the top graphene contact and the CrI$_3$ barrier, its effect on the bottom graphene contact is negligible. Similarly, the effect of the bottom gate on the top graphene contact is also negligible. This is clearly seen in our devices (Supplementary Fig. S3 and S7) and we will focus on the top gate below, which influences only the Fermi level of the top graphene contact.

The gate dependence of the tunnel conductance shows three distinct regions: *p-p*, *p-i* and *p-n* (denoting the doping type in the bottom and top bilayer graphene contact, respectively) from left to right in Fig. 1b. The insets illustrate the band alignments of the two graphene contacts and the tunnel barrier. In the *p-i* region, the tunnel junction is off since the Fermi level of the bottom graphene contact falls inside the gap of the top graphene contact and no state is available for resonant tunneling in the zero-bias limit. In the *p-n* region, when the junction is forward biased, electrons tunnel from the conduction band on the *n*-side to the unoccupied states in the valence band on the *p*-side through the CrI$_3$ barrier (upper panel, Fig. 1c). The tunnel current *I* reaches a peak when the Fermi level on the *n*-side aligns with the valence band maximum on the *p*-side. Further increase in the bias voltage $V_b$ decreases the tunnel current, which is more clearly seen in the negative differential conductance d*I*/d$V_b$ (dip 1 in lower panel, Fig. 1c and inset). The negative differential conductance is the key signature of a *p-n* tunnel diode [24, 25, 28]. In the *p-p* region, negative differential conductance disappears as expected. Instead, there appears a kink in the tunnel current, or equivalently, a dip in the differential conductance on each side of zero bias (Fig. 1d). The dip under reversed biasing (dip 3) corresponds to the Fermi level of the bottom graphene contact inside the gap of the top graphene contact. It shifts towards higher reversed bias voltages for lower Fermi energies (i.e. higher hole doping) in the top graphene contact. On the other hand, the dip under forward biasing (dip 2) depends weakly on $V_{tg}$ since it corresponds to the Fermi level of the top graphene contact inside the gap of the bottom graphene contact, whose Fermi level is hardly affected by top gating. The behavior of dip 2 and 3 is switched when a bottom gate voltage $V_{bg}$ is applied instead of the top gate voltage (Fig. 1e), which is consistent with the picture given above and confirms that the two gates are symmetric. The dispersion of the dip positions (1-3) with $V_{tg}$ is summarized in Fig. 1f (symbols) and is in good agreement with the resonant tunneling model (solid lines, details presented in Supplementary Sect. 5).

Equipped with the above basic understanding of the operation of our TFET, we now examine its TMR in Fig. 2a. The upper panel compares the gate dependence of the tunnel conductance *G* under 0 T and 1 T and shows in general larger *G*'s at 1 T. Figure 2c and 2f are the detailed magnetic-field dependence of *G* for a representative gate voltage from the *p-p* and *p-n* region, respectively. The tunnel conductance exhibits a sharp jump around 0.5 T, which is associated with the spin-flip transition in bilayer CrI$_3$ (See Supplementary Fig. S4 for independent verification by magnetic circular dichroism). The transition field depends on gate voltage [7, 8], which is a key property to realize the spin TFET as we discuss below. The magnetic hysteresis reflects the first-order nature of the spin-flip transition [29]. We calculate the TMR $= \frac{G_{FM} - G_{AFM}}{G_{AFM}}$ using the conductance at 0 T and 1 T for $G_{AFM}$ and $G_{FM}$, respectively, in the lower panel of Fig. 2a. It is ~ 80% in the *p-p* and *p-n* regions and is weakly gate dependent. The TMR here is governed by the spin filtering efficiency of CrI$_3$ (the ratio of the transmission probability of electrons with spins parallel to antiparallel to magnetization in each CrI$_3$ monolayer), which can be estimated to be ~ 3.3 and is in good agreement with the reported values [9, 10]. What's more interesting, however, is the behavior of TMR near the boundaries of the off state of the TFET. An enhancement is observed at the *p-p* and *p-i* boundary (Fig. 2d). The TMR even becomes negative with large amplitude at the *p-i* and *p-n* boundary, *i.e.* $G_{FM} < G_{AFM}$ (Fig. 2e). A careful examination of Fig. 2a indicates that the gate dependence of $G_{FM}$ is shifted from that of $G_{AFM}$ towards larger $V_{tg}$'s. Because of the ambipolar transistor behavior of the TFET, this shift leads to an enhancement of the TMR and a sign change at the boundaries of the off state. Such a behavior has been observed in all devices with the same structure studied in this work although the detailed values of the TMR vary from device to device. Figure 2b is the result for device 2 showing peak TMR values of ~ 1,000% and -10,000% at the boundaries of the off state.

To understand the shift between the gate dependence of $G_{AFM}$ and $G_{FM}$, we have performed detailed bias dependence measurements of the tunnel conductance under varying magnetic fields. Figure 3a and 3b are the differential conductance d$I$/d$V_b$ as a function of bias voltage for a representative *p-n* and *p-p* junction, respectively. The curves under different magnetic fields ranging from -1 T to 1 T with a step of 0.1 T are vertically displaced for clarity. We track the position in bias voltage of differential conductance dip 1 - 3 as a function of magnetic field (Fig. 3c - 3e). All three dip positions show a clear jump across the spin-flip transition of bilayer CrI$_3$ with the ferromagnetic state occurring at a larger absolute bias. This jump can be understood as a downshift of the Fermi energy in the graphene contacts as a result of the spin-flip transition. For instance, the negative differential conductance for the *p-n* junction (dip 1) upshifts by ~ 20 mV across the spin-flip transition (Fig. 3c). It corresponds to a downshift of the graphene Fermi level by ~ 20 meV (Fig. 2f). The downshift of the Fermi level is equivalent to having a higher unintentional hole doping density, thus shifting the gate dependence of $G_{FM}$ towards larger $V_{tg}$'s. Figure 3c - 3e also show the gate dependence of the height of each jump and the absolute position of the dips, which are discussed in Supplementary Sect. 4. A plausible mechanism for the Fermi level shift is a magnetic order-dependent band structure in CrI$_3$ [30]. This can change the built-in electric field as well as charge transfer at the graphene/CrI$_3$ interfaces. Such a picture is supported by the observed TMR with $G_{FM} > G_{AFM}$ away from the boundaries of the off state because spin filtering can also be viewed as a reduction in the tunnel barrier height for ferromagnetic CrI$_3$ [9-12]. Future theoretical studies with accurate band structure calculations are needed for a quantitative description of our observation. Nevertheless, we summarize in Fig. 2c-f schematically the graphene band alignments across the spin-flip transition and the corresponding behavior of TMR. In the *p-p* and *p-n* regions (Fig. 2c, 2f), TMR is originated from spin filtering, as reported by Ref. [9-12]. Near the boundaries of the *p-i* region (Fig. 2d, 2e), the significantly modified TMR is caused by a magnetic transition-induced Fermi level shift in the graphene contacts.

Finally we demonstrate the spin transistor action based on our TFET devices, *i.e.* control of the tunnel conductance $G$ by electrically switching the magnetic order in the tunnel barrier. To minimize the trivial conductance change due to electrostatic doping, we choose a range of gate voltage from the *p-p* region ($V_{tg} < 0$ V) where $G$ is weakly gate dependent in the absence of magnetism. Figure 4a shows the gate dependence of $G$ under a constant magnetic bias (-0.76 T), which is slightly above the spin-flip transition of bilayer CrI$_3$ under $V_{tg} = 0$ V. The gate voltage can repeatedly alter the TFET between a high (red symbols) and a low (black symbols) conductance state that have been achieved in Fig. 2 under 1 T and 0 T with the CrI$_3$ barrier in the ferromagnetic and antiferromagnetic state, respectively. A relative change of ~ 35% in the tunnel conductance has been achieved through gating and with a large hysteresis. The switching is originated from a gate-dependent spin-flip transition field, as discussed above. The relative change in the tunnel conductance can be significantly enhanced by using thicker tunnel barriers on the expense of the tunnel conductance. Figure 4b shows the action of a spin-TFET with 4-layer CrI$_3$ barrier under a constant magnetic bias of -1.77 T (Supplementary Sect. 7 for additional characterizations of this device). Nearly 400% change in the tunnel conductance is achieved while the tunnel conductance is an order of magnitude lower than that in the spin-TFET with bilayer CrI$_3$ (Fig. 4a). In conclusion, we have demonstrated efficient electrical writing and readout of the magnetic states in a single spin-TFET device based on graphene-CrI$_3$ heterostructures. The device concept developed in this study will open up a new avenue for exploring power-efficient non-volatile memory devices based on electric-field controlled magnetism [31].

**Methods**
**Device fabrication.** Van der Waals heterostructures of CrI$_3$, hexagonal boron nitride (hBN) and graphene were fabricated by the layer-by-layer dry transfer method [32]. Images of sample devices are shown in Supplementary Fig. S1. In this process, atomically thin samples of CrI$_3$, hBN and graphene were first exfoliated from their bulk crystals (HQ graphene) onto silicon substrates covered with a 300-nm thermal oxide layer. The thickness of thin flakes was initially estimated from their optical reflectance contrast on silicon substrates and later verified by the atomic force microscopy for hBN and few-layer graphene samples, and Raman spectroscopy for bilayer graphene. Bilayer CrI$_3$ was confirmed by the magnetization measurement under an out-of-plane magnetic field. The selected thin flakes of appropriate thickness and geometry were then picked up one-by-one by a stamp consisting of a thin layer of polycarbonate (PC) on polydimethylsiloxane (PDMS). The complete stack was then deposited onto substrates with pre-patterned Au electrodes. The residual PC on the device surface was dissolved in chloroform before measurements. Since atomically thin CrI$_3$ is extremely air sensitive, it was handled in a nitrogen gas environment with less than one-part-per-million oxygen and moisture inside a glovebox and was removed from the glovebox only after being fully encapsulated.

**Electrical characterization.** The device characterization was carried out in an Attocube closed-cycle cryostat (attoDry1000) down to 4 K and up to 2.5 T. The tunnel conductance of the graphene/CrI$_3$ tunnel junctions was measured by biasing CrI$_3$ using the bilayer graphene source and drain contacts (Fig. 1a). The bias voltage was kept in the low bias regime (4 mV in amplitude). For the magnetic field and gate voltage dependence studies of the tunnel conductance, the bias was modulated at a relatively low frequency of about 17 Hz and the resultant current was measured with a lock-in amplifier to reduce noise. The tunnel conductance was calculated as the ratio of current to bias. The bias dependence of the tunnel current was studied by applying a dc bias voltage using a sourcemeter and measuring the resultant dc current using a current preamplifier. The differential conductance as a function of bias voltage was obtained via numerical differentiation of the bias dependence of the tunnel current.

**Magnetic circular dichroism.** The magnetization of CrI$_3$ was characterized by magnetic circular dichroism (MCD) using a HeNe laser at 633 nm. An objective of numerical aperture 0.8 was used to focus the excitation beam to a sub-micron spot size on the devices. The optical power on the devices was limited to 5 μW. The reflected light from the devices was collected by the same objective and detected by a photodiode. The helicity of the optical excitation was modulated between left and right by a photoelastic modulator at 50.1 kHz. The MCD was determined as the ratio of the ac component (measured by a lock-in amplifier) to the dc component (measured by a multimeter) of the reflected light intensity.

**Figures and figure captions**

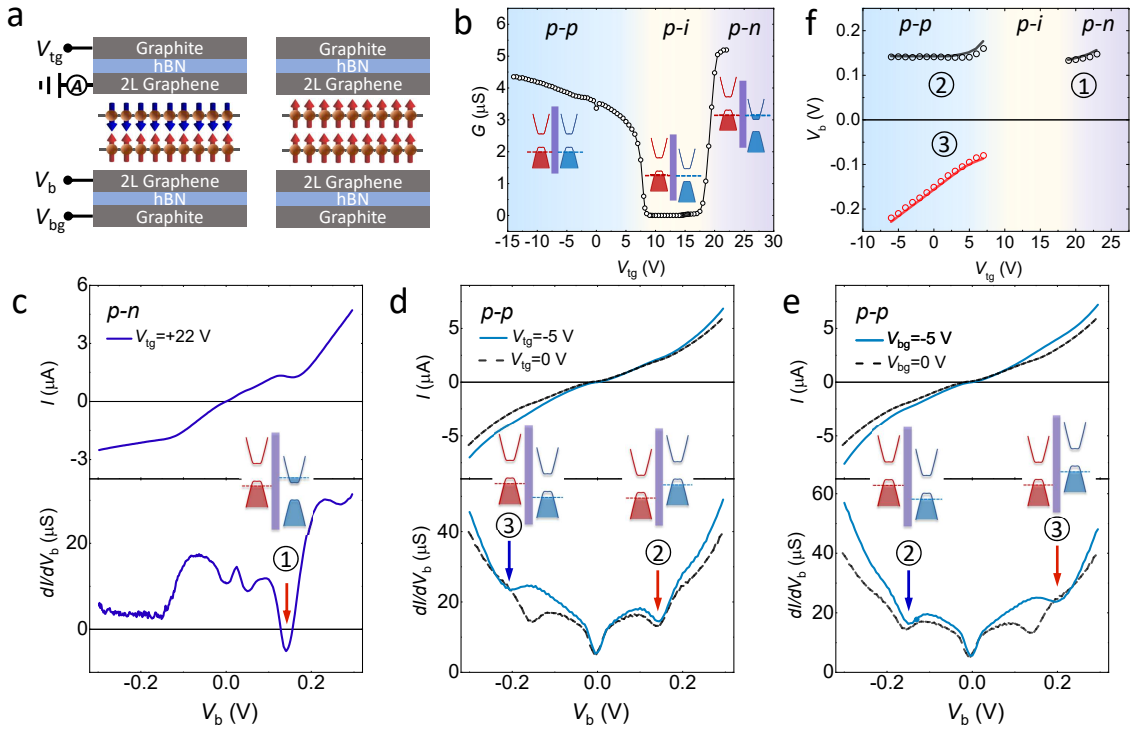

**Figure 1 | Tunnel field-effect transistor (TFET) based on graphene/CrI$_3$ heterostructures. a**, Operational principle of a spin TFET based on gate-controlled spin-flip transition in bilayer CrI$_3$ and spin filtering in the tunnel junction. Arrows indicate the spin orientation in CrI$_3$ layers. The left and right panels correspond to a low and a high tunnel conductance state, respectively. **b**, Gate dependence of tunnel conductance shows *p-p*, *p-i* and *p-n* regions. The feature at zero gate voltage is due to the resistance of the graphene contact in series with the tunnel junction. **c**, **d**, Bias dependence of tunnel current (upper panel) and differential conductance (lower panel) for representative *p-n* (**c**) and *p-p* (**d**) junctions. **e**, Same as **d** under a bottom instead of a top gate voltage. **f**, Gate dependence of bias voltages that correspond to differential conductance dip 1 - 3 as defined by arrows in **c** and **d**. The insets in **b** - **e** show the band alignments of the bottom (red) and top (blue) bilayer graphene contacts and the CrI$_3$ barrier (purple). Shaded areas are filled states.

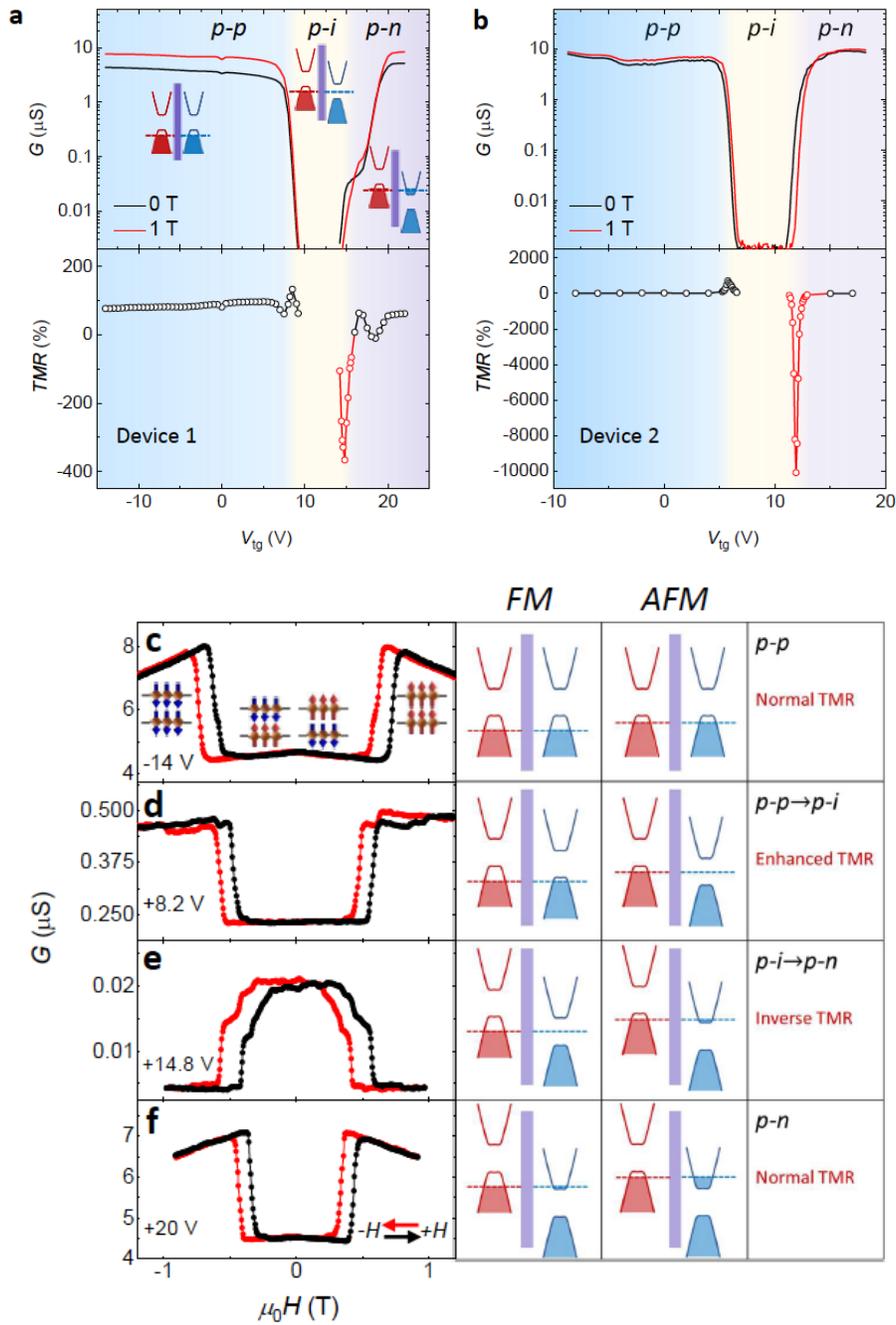

**Figure 2 | Tunable tunnel magnetoresistance (TMR). a**, Gate dependence of tunnel conductance under an out-of-plane magnetic field of 0 T and 1 T (upper panel), and TMR (lower panel) calculated from the values in the upper panel for device 1. The black and red symbols denote, respectively, the positive and negative values for TMR. The insets show the band alignments of the *p-p*, *p-i* and *p-n* tunnel junctions. **b**, Same as **a** for device 2. **c – f**, The left column is the magnetic-field dependence of tunnel conductance at varying top gates; middle two columns, the band alignments of the junction when CrI$_3$ is in the ferromagnetic (FM) and antiferromagnetic (AFM) state; right column, the behavior of TMR in the *p-p* (**c**), boundaries of *p-i* (**d**, **e**), and *p-n* (**f**) region.

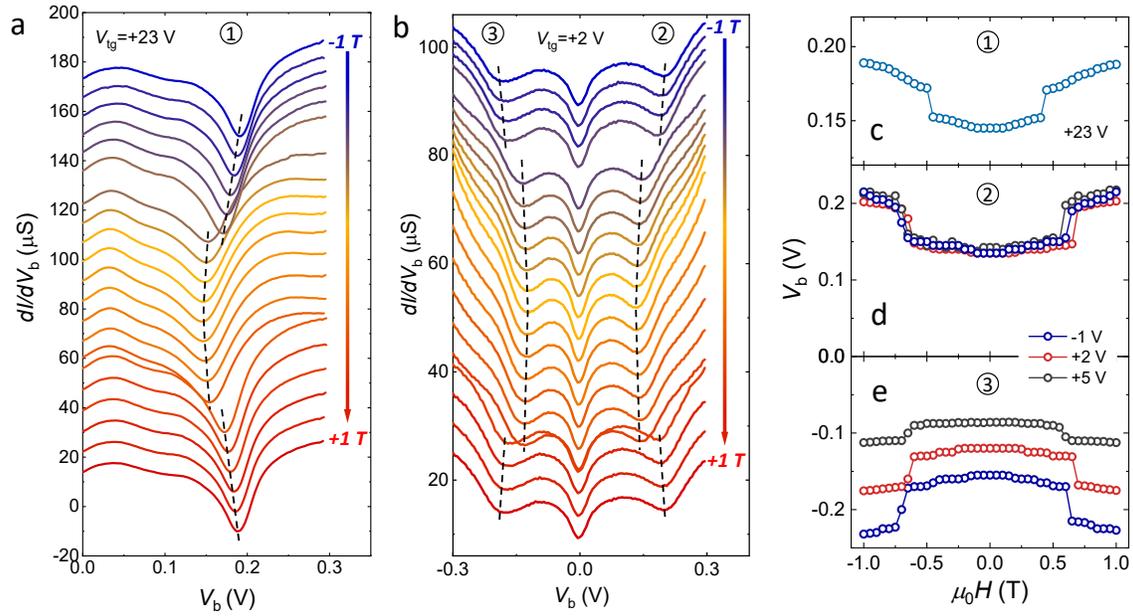

**Figure 3 | Magnetic transition-induced band shift. a**, **b**, Bias dependence of differential tunnel conductance of device 1 for a representative *p-n* (**a**) and *p-p* (**b**) junction. Results for magnetic field varying from 1 T to -1 T with a step size of 0.1 T are vertically displaced by the same amount for clarity. The dashed lines are guides to the eye for the evolution of differential conductance dip 1 (**a**) and dip 2 and 3 (**b**) as defined in Fig. 1c and 1d. **c - e**, Magnetic-field dependence of bias voltage corresponding to dip 1 (**c**), 2 (**d**) and 3 (**e**) for representative top gate voltages. The magnetic field was scanned from -1 T to 1 T.

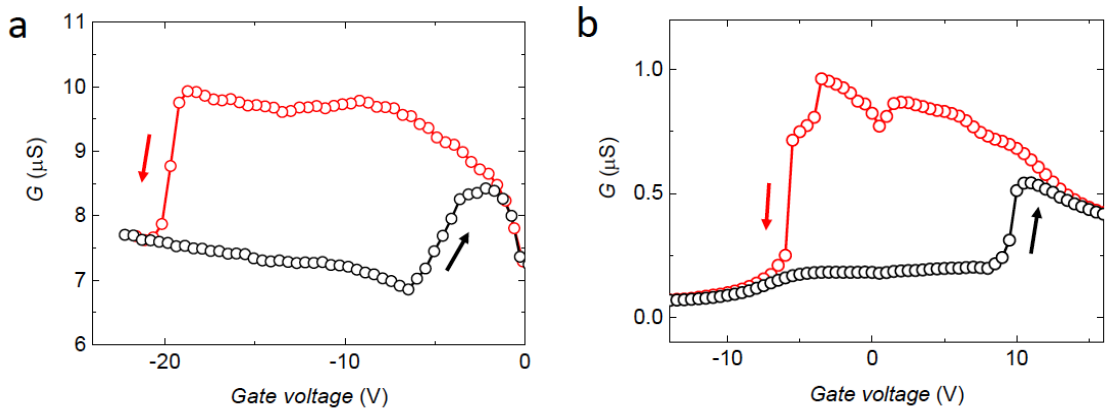

**Figure 4 | Spin TFET action. a**, Tunnel conductance of a TFET with a bilayer CrI$_3$ tunnel barrier is repeatedly switched by gating under a constant magnetic bias of -0.76 T. The top and bottom gate voltages are identical and their sum is shown in the horizontal axis. Black and red symbols correspond to measurements while sweeping the gate voltage forward and backward, respectively. A change of ~ 35% is seen in the tunnel conductance. **b**, The same as in **a** for a TFET with a four-layer CrI$_3$ tunnel barrier under a constant magnetic bias of -1.77 T. A change of ~ 400% is seen in the tunnel conductance.